\documentclass[fleqn,10pt]{wlscirep}
\usepackage[utf8]{inputenc}
\usepackage[T1]{fontenc}

\usepackage[table]{xcolor}
\usepackage{array}

\definecolor{headerblue}{HTML}{002060}    
\definecolor{lightgreen}{HTML}{E2EFDA}    
\definecolor{lightred}{HTML}{FCE4D6}     

\newcommand{\cell}[2]{%
  \begin{tabular}{@{}c@{}}
    #1 \\ #2
  \end{tabular}%
}

\newcommand{\gcell}[2]{%
  \cellcolor{lightgreen}%
  \begin{tabular}{@{}c@{}}
    #1 \\ #2
  \end{tabular}%
}

\newcommand{\rcell}[2]{%
  \cellcolor{lightred}%
  \begin{tabular}{@{}c@{}}
    #1 \\ #2
  \end{tabular}%
}

\newcommand{\headercell}[2]{%
  \begin{tabular}{@{}c@{}}
    \textcolor{white}{\textbf{#1}} \\
    \textcolor{white}{\textbf{#2}}
  \end{tabular}%
}

\title{The Power of Altruism in Sticker Economics: Generosity Minimizes Collective Costs and Overprotective Norms Fuel Inefficiency}

\author[1]{Luana Ferraz Alvarenga}
\author[2]{Caetano Alvarenga Costa}
\author[1,*]{César Rennó-Costa}
\affil[1]{Universidade Federal do Rio Grande do Norte, Instituto Metrópole Digital, Natal, Brazil}
\affil[2]{Apoema Escola Dialógica, Pitanga, Natal, Brazil}

\affil[*]{cesar@imd.ufrn.br}

\begin{abstract}
Collecting the FIFA World Cup sticker album presents a classic public-goods and collective-action dilemma, in which completing a collection on one's own is highly inefficient. To evaluate how localized community norms shape collective efficiency, we use agent-based modeling and Monte Carlo simulations, parameterized with empirical field observations from exchange meetups in Natal, Brazil. Reflecting the tournament's recent expansion, the Panini 2026 album features 980 individual stickers, including 68 metallic specials. We contrast a standard baseline economy (1:2 special-to-normal exchange ratio) with an overprotective, strict strategy (exclusive special-for-special trading) and an altruistic, generous strategy (in which advanced players surrender needed duplicates to assist peers).  Our findings reveal that overprotective rules trap liquidity and drive network-wide inefficiency. The strict strategy increases median completion costs by 10 packs and severely penalizes the least fortunate 5\% of collectors, adding 20 packs in large cities and 30 in small communities. Conversely, widespread generosity optimizes network liquidity and dramatically compresses the long tail of bad luck. Introducing the generous strategy reduces required purchases for the 5th percentile by 90 packs in large-scale configurations and 130 packs in smaller clusters. Furthermore, widespread altruism triggers a strong functional coupling that effectively synchronizes completion rates across the network. This study demonstrates that while rigid, protective norms degrade collective welfare, generosity successfully mitigates pack-draw variance, transforming an expensive, isolated hobby into a resilient, highly efficient public good.

\end{abstract}
\begin{document}

\flushbottom
\maketitle
%
%
\thispagestyle{empty}

\section*{Introduction}

As the FIFA World Cup approaches, so does the frenzy of collecting the tournament's sticker album. This creates a fascinating challenge for football fanatics, especially those with young kids, who find themselves trying to explain the true value of completing the collection to the uninitiated. To an outsider, the cost of buying every single sticker alone seems outstanding, if not completely prohibitive. However, the fundamental point is that collecting an album is, at its core, a social event. Within football fandom culture, such consumption practices serve key communal functions, facilitating identity, shared rituals, and socialization \cite{AyKaygan2021}. Viewed through a sociological and game-theoretic lens, completing an album alone is highly inefficient because a single member's total contribution cost far exceeds their individual benefit. This imbalance mirrors a structural friction between short-term individual optimization and long-term collective welfare. Ultimately, the perceived value is anchored far more in participating in a collective endeavor than in actually owning the final product, which is not genuinely rare. The true worth of the album emerges from the complex behavioral dynamics of public goods and collective action dilemmas, where voluntary participation often shifts away from narrow self-interest toward conditional cooperation based on peer interactions \cite{KatuscakMiklanek2022}. This coordinated architecture of shared human connection, social norms, and mutual reward transforms what would otherwise be an expensive, isolated hobby into a resilient public good.

As a social phenomenon, this hobby offers an interesting context for evaluating individual behavior within a collective setting. The core debate is whether individual norms and decisions benefit the collective through cooperation or whether they protect the individual within a more competitive framework. Crucially, different sets of rules can have independent impacts on individual protection and collective costs. For instance, exchanging duplicates is mutually beneficial. This behavior functions as an informal networking strategy where agents engage in reciprocal cooperation to satisfy individual constraints while creating positive spillovers for the broader community \cite{pyka_informal_2000}. It allows collectors to complete the album by buying fewer sticker packs, demonstrating that traditional face-to-face peer-to-peer sharing dramatically reduces individual acquisition costs and resource waste while simultaneously fostering social connectivity \cite{Hansmann03102023}.

On the other hand, some rules create a trade-off where the group benefits at the expense of the individual, or the individual benefits at the expense of the group. For example, recent editions of the album introduced special cards with a metallic finish, and although there is no evidence that they are actually rarer, their high perceived value led the community to develop specific, rigid rules for trading them. This behavior highlights how perceived product scarcity and subjective attributes radically distort a consumer's information processing and choice evaluations, independent of actual market supply \cite{Hamiltonetal2018}. Furthermore, a natural asymmetry always exists regarding how close each player is to finishing their collection. In many trading situations, one player will inevitably need fewer duplicates than the other. Even though a single exchange is technically more valuable to the player who is closer to completion, community trading norms rarely accommodate these asymmetric conditions, opting for a strict one-for-one rule instead. This institutional rigidity demonstrates how informal peer groups often reject mathematically optimal, variable benefit-distribution schemes in favor of simple, egalitarian cost-sharing baselines to avoid negotiation friction \cite{Parrachinoetal2006}.

In this study, we use computer modeling and a Monte Carlo approach, similar to those used in epidemiology \cite{lopes_measuring_2022} and neuroscience \cite{rennocosta_place_2017}, to evaluate how variations in trading rules impact both individual and collective gains for the Panini 2026 World Cup album. Reflecting the tournament's recent expansion in the number of participating teams \cite{renno-costa_double-elimination_2023}, the album features a notable increase in size compared to previous editions, totaling 980 individual stickers, including 68 with a metallic finish, and is sold in packs of 7. While the album also includes brand-specific and extra stickers, they add a layer of complexity that falls outside the scope of this analysis. Specifically, we evaluate the distinct rules governing the exchange of regular versus metallic stickers, as well as the behavioral dynamics of asymmetric trading. Ultimately, our goal is to better understand this social phenomenon and shed light on how localized community norms dictate the efficiency of collective action.

\section*{Results}

\begin{figure}
    \centering
    \includegraphics[width=1\linewidth]{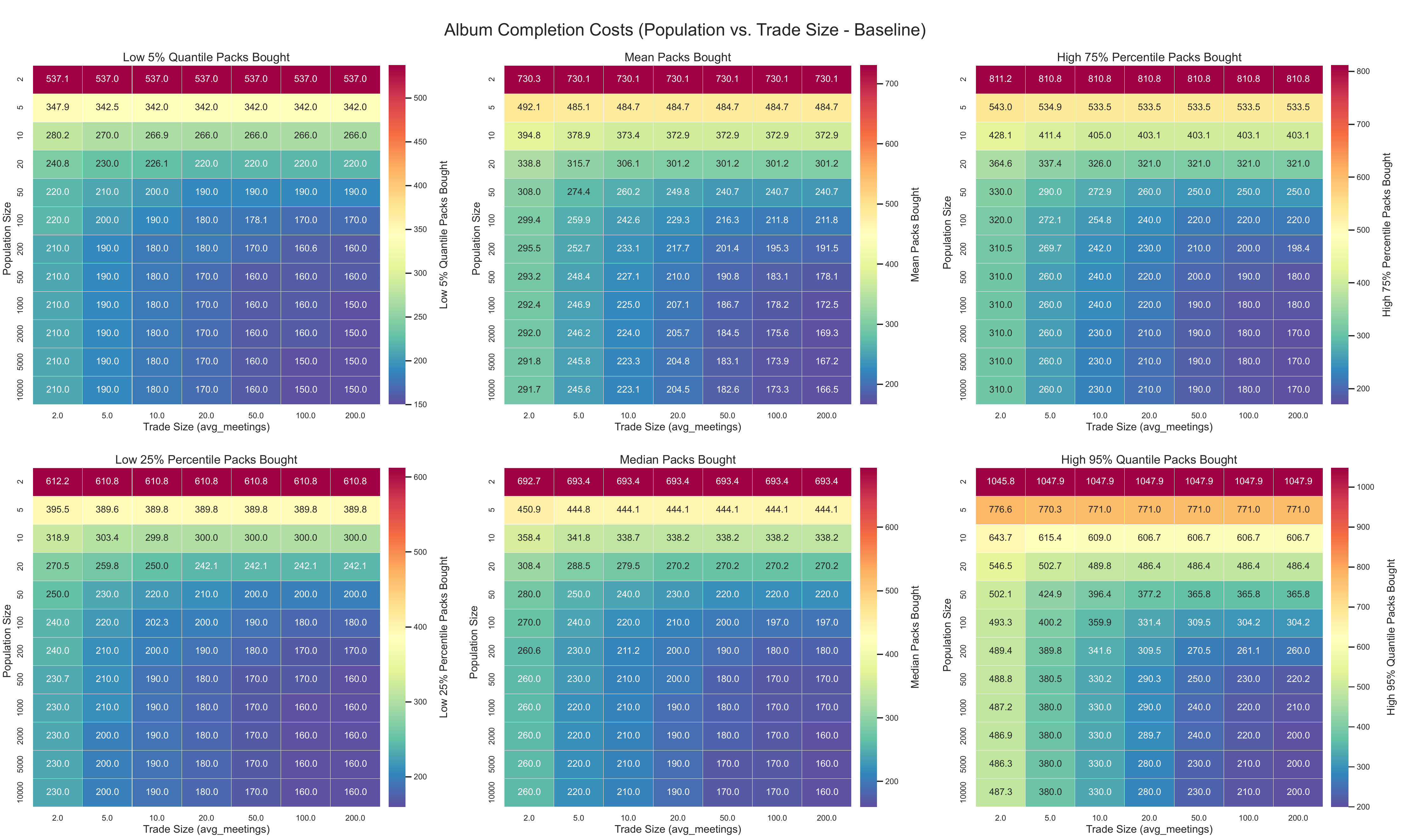}
    \caption{\textbf{Impact of population size and trade frequency on album completion costs.} Heatmaps display the distribution of the total number of sticker packs required to complete the album as a function of the total number of players in the simulation (vertical axis, ranging from 2 to 10,000 players) and the average number of exchange meetings per player (horizontal axis, ranging from 2 to 200 meetings). To capture the variance in individual outcomes across the aggregated simulation runs, six statistical metrics are presented across the subpanels: Median (center), Mean (bottom center), 25th Percentile (bottom left), 75th Percentile (bottom right), 5th Quantile (top left), and 95th Quantile (top right). The color gradients indicate the volume of packs bought, with warmer hues (reds/oranges) corresponding to higher individual completion costs and cooler hues (blues/greens) denoting greater collective efficiency and lower overall costs.}
    \label{fig:heatmap}
\end{figure}

To initially evaluate the impact of network scale on the simulation dynamics, we executed the baseline model using varying population sizes and adjusted the number of exchange encounters permitted between sticker pack acquisitions (Figure \ref{fig:heatmap}). As anticipated, expanding the player pool dramatically reduced the total number of packs required to complete the album. In a minimal baseline simulation involving only two mutually trading players, the average requirement exceeded 740 packs, with the least fortunate 25\% of participants needing over 800 packs. This represents more than a five-fold increase over the theoretical minimum of 140 packs—the absolute baseline required if a player could perfectly exchange every duplicate. Conversely, in a large-scale population of 1,000 players who conduct over 200 exchanges, the average cost plummeted to 166 packs per player, falling to less than 15\% above the theoretical optimum.

\begin{figure}
    \centering
    \includegraphics[width=1\linewidth]{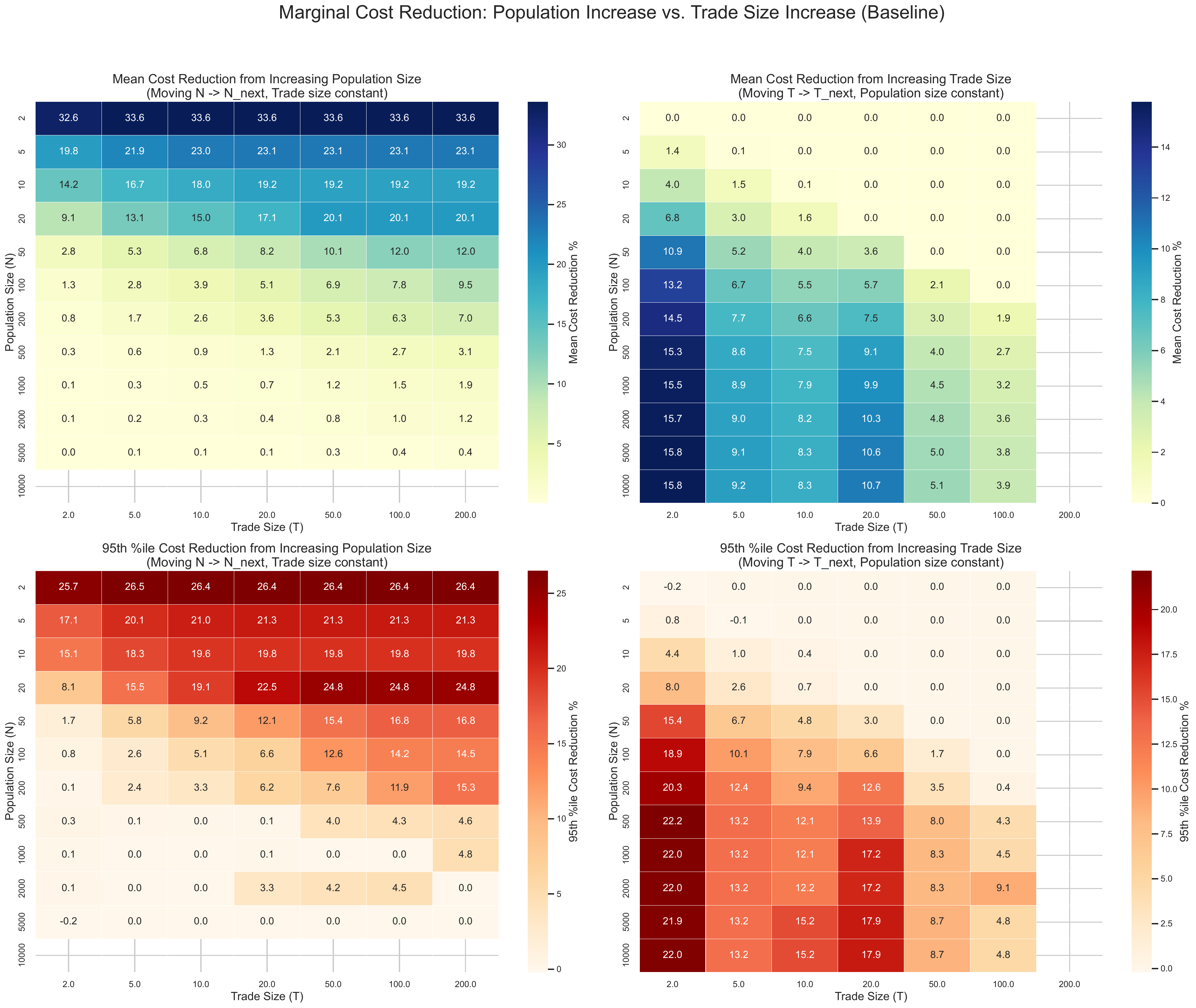}
    \caption{\textbf{Marginal cost reduction in album completion driven by incremental increases in population versus trade size}. The heatmaps illustrate the relative percentage decrease in the number of sticker packs required when transitioning to the next simulated tier of either population size (N) or trade size (T), holding the other variable constant. The top row (blue-green gradient) displays the marginal reduction in the mean completion cost across the population, while the bottom row (orange-red gradient) shows the marginal reduction for the 95th percentile (representing the cost threshold for the least lucky 5\% of collectors). The left column isolates the effect of population growth, calculating the efficiency gained by moving from the current population tier to the next ($N -> N_{next}$) at a fixed trade size. Conversely, the right column isolates the effect of increased trading frequency, showing the cost reduction when moving to the next trade size tier ($T -> T_{next}$) within a fixed population. Cell values represent the percentage drop in required packs; darker shades indicate critical thresholds where incremental expansions in either community size or trade limits yield the most substantial efficiency gains.
}
    \label{fig:bottleneck}
\end{figure}

While population size proved a major driver in reducing completion costs, the overall efficacy was clearly bottlenecked by the frequency of trades executed among the players (Figure \ref{fig:bottleneck}). Our marginal cost reduction analysis reveals that simply adding more participants yields diminishing returns if the interaction bandwidth remains restricted. For instance, in large networks, increasing the population size without a proportional expansion of trade opportunities reaches a saturation point, where excess duplicates remain trapped within localized clusters. Conversely, amplifying the average number of exchange meetings within a stable population unlocks substantial efficiency gains by mobilizing these stagnant assets. This interaction bottleneck is particularly severe for the least fortunate collectors—those at the 95th percentile. For these individuals, the variance of poor sticker pack draws can only be mitigated through high-frequency trading, which allows them to fully exploit the network's broader liquidity. Ultimately, achieving near-optimal collective efficiency requires a structural synergy between community scale and interaction capacity; a massive pool of collective duplicates is only beneficial if the social architecture permits those stickers to flow freely to the individuals who need them most.

Having established the relationship between population size, trading frequency, and the overall cost of album completion, we shifted our focus to the impact of localized trading rules. For this phase of the analysis, we defined two distinct network topologies: a small-city configuration comprising a cluster of 100 players executing an average of 10 trades per round, and a large-city configuration comprising 1,000 players executing 20 trades per round.

\begin{figure}
    \centering
    \includegraphics[width=1\linewidth]{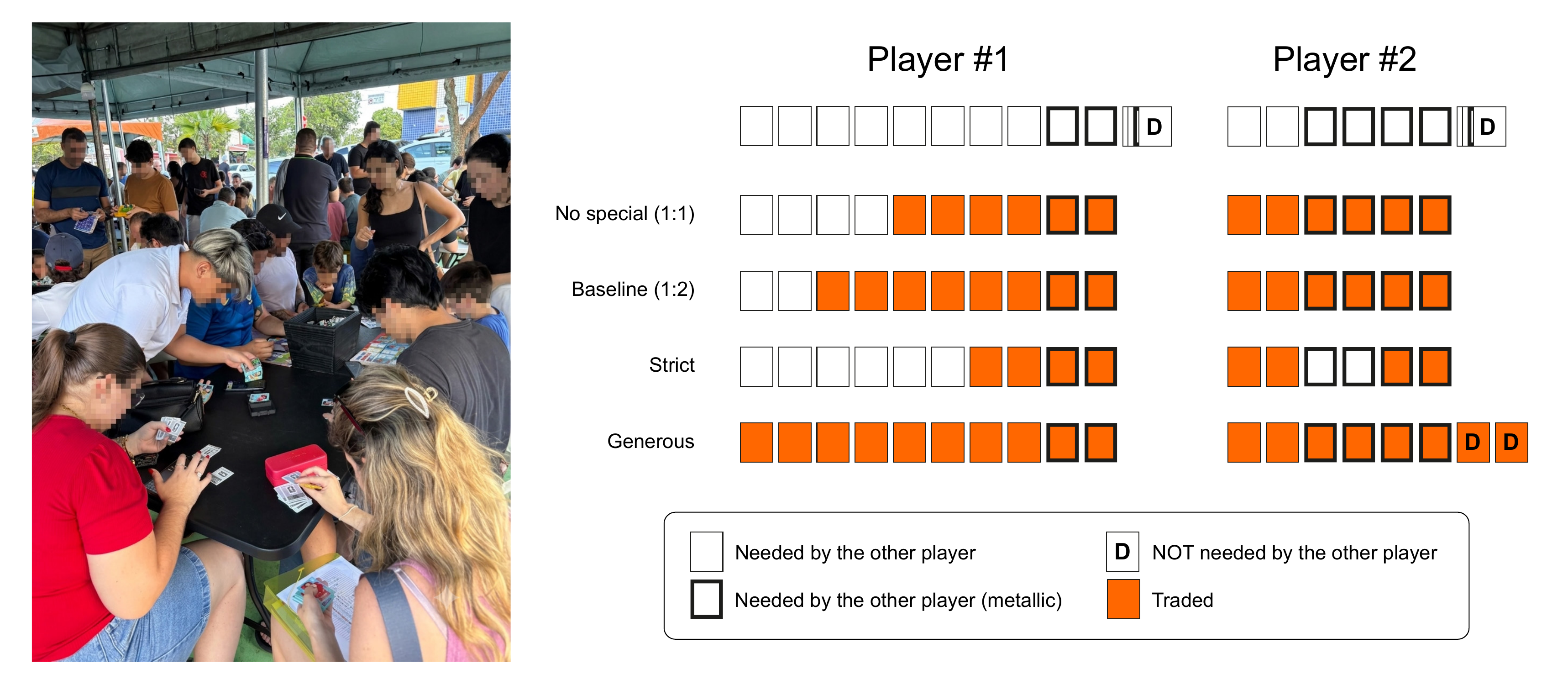}

    \caption{\textbf{Empirical field observations of a sticker-trading event and schematic frameworks of localized community trading norms.} The left panel captures direct empirical field observations of a local sticker exchange meetup in Natal, Brazil, highlighting the organic social mixing, dense interaction bandwidth, and multi-generational dynamics of fans engaging in collective action. The right panel provides a schematic strategic framework mapping these human behaviors into four distinct agent-based modeling scenarios between Player 1 and Player 2: No special (1:1) represents a uniform approach where collectors ignore the sticker finish to execute standard one-for-one trades based strictly on mutual need; Baseline (1:2) reflects the prevailing community norm where regular stickers are traded normally but metallic specials command a one-to-two exchange ratio against regular ones; Strict models a rigid, protectionist segregation rule that restricts metallic stickers exclusively to special-for-special trading and entirely prohibits cross-category exchanges; and Generous outlines an altruistic strategy targeting behavioral asymmetry where an advanced collector surrenders needed duplicates to a peer, accepting unneeded assets simply to satisfy point-value inventory balancing. The accompanying variables define the exact operational mechanics of these rules, with duplicate regular/metallic stickers where \textit{Needed by the other player} identifies assets directly addressing a partner's collection gaps, \textit{NOT needed by the other player} denotes arbitrary duplicate fillers used as transactional ballast for inventory balancing, \textit{Needed by the other player (metallic)} indicates required special-finish assets, and \textit{Traded} marks the formal execution and structural resolution of the exchange pair. }
    \label{fig:trading}
\end{figure}

Within these environments, we modeled empirical trading strategies based on field observations at local sticker exchange meetups in Natal, Brazil, mapping these human behaviors into structural agent-based modeling scenarios \ref{fig:trading}. The primary set of rules we evaluated governed the exchange dynamics between metallic and regular (non-metallic) stickers. At these physical meetups, the prevailing community norm dictated a 1:2 exchange ratio, in which a single metallic sticker was valued at two regular stickers. However, we also observed—and subsequently incorporated into our models—two alternative behaviors: a strict segregation rule where metallic stickers were exclusively traded for other metallic stickers (and regular for regular), and a no special approach where collectors ignored the sticker finish entirely, executing all trades at a standard one-to-one ratio regardless of type.

\begin{figure}
    \centering
    \includegraphics[width=1\linewidth]{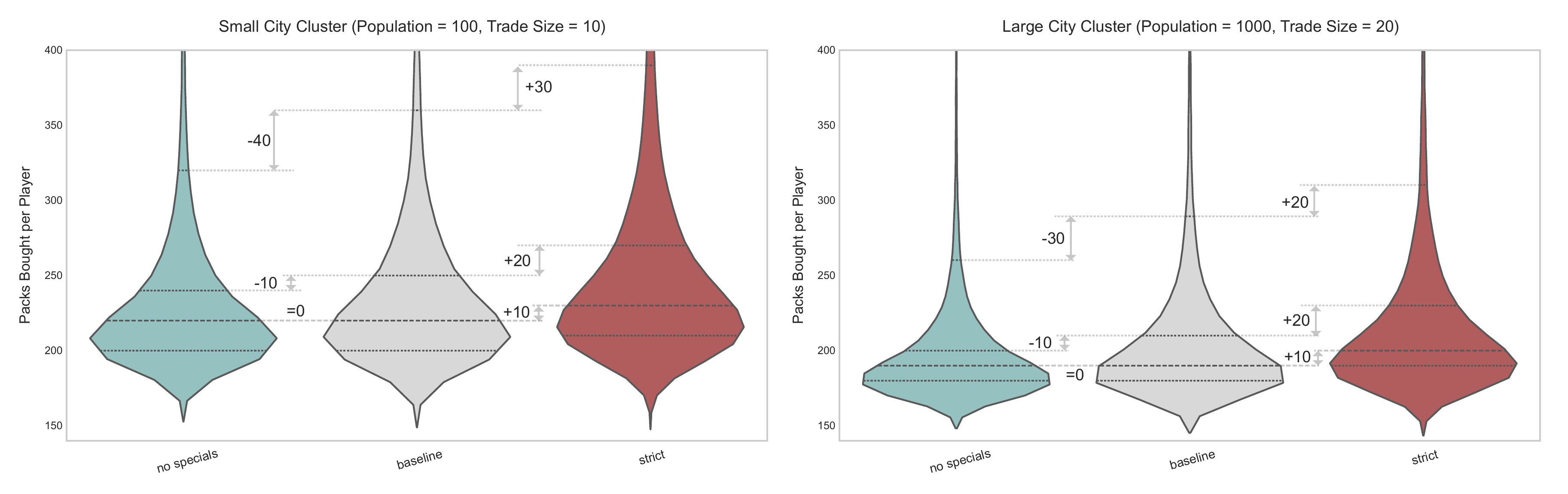}
    \caption{Impact of localized trading rules on the distribution of album completion costs. Violin plots illustrate the distribution of the total sticker packs required per player across three distinct trading strategies: no specials (ignoring sticker finish, trading all 1:1), baseline (standard 1:2 exchange ratio for metallic stickers), and strict (exclusive special-for-special exchange). The analysis contrasts two network topologies: a Small City Cluster (population = 100, trade size = 10; left panel) and a Large City Cluster (population = 1,000, trade size = 20; right panel). The width of each violin represents the density of players finishing at that specific cost threshold. Horizontal dashed lines within each violin denote the 25th, 50th (median), and 75th percentiles. Dotted horizontal lines and numeric annotations highlight the absolute marginal differences (in packs) of the median and the 95th percentile (the least lucky 5\% of collectors) when compared to the baseline condition. The strict strategy consistently shifts the distribution upward, worsening costs, while the no specials strategy reduces the long-tail cost burden for the least fortunate players without affecting the median.}
    \label{fig:strict_pop}
\end{figure}

In an initial analysis, we evaluated the global impact of each strategy on the primary objective of trading: minimizing the collective cost of completing the albums (Figure \ref{fig:strict_pop}). The strict strategy negatively affected all evaluated scenarios, including both the small-city and large-city configurations. This impact manifested in the median cost, requiring an additional 10 packs in both conditions, and severely affected the least fortunate 5th percentile of players, adding 20 extra packs in the large city and 30 extra packs in the small city. Interestingly, adopting a strategy that ignores special finishes had zero effect on the median cost in either city. However, this approach yielded a noticeable benefit for the 5th percentile, reducing the required number of packs by 30 in both environments. Comparing the highly efficient uniform strategy to the highly inefficient strict trading method reveals a stark contrast for the least lucky players. The difference amounted to 50 packs for the large city and an astounding 60 packs for the small city, representing over 40 percent of the requirements for an optimal situation. A comprehensive percentile breakdown of these baseline, strict, and uniform conditions across both network scales is summarized in Table \ref{tab:tables_small} and Table \ref{tab:tables_large}.

\begin{center}
    \begin{tabular}{|l|c|c|c|c|c|c|}
        \hline
        \rowcolor{headerblue}
        \headercell{Condition}{} & 
        \headercell{5\%}{(abs / diff)} & 
        \headercell{25\%}{(abs / diff)} & 
        \headercell{Median}{(abs / diff)} & 
        \headercell{Mean}{(abs / diff)} & 
        \headercell{75\%}{(abs / diff)} & 
        \headercell{95\%}{(abs / diff)} \\ \hline
        
        generous-nospecial & \cell{190.0}{(+0.0)} & \gcell{190.0}{(-10.0)} & \gcell{200.0}{(-20.0)} & \gcell{203.9}{(-38.6)} & \gcell{210.0}{(-50.0)} & \gcell{230.0}{(-130.0)} \\ \hline
        
        generous-baseline & \cell{190.0}{(+0.0)} & \cell{200.0}{(+0.0)} & \gcell{200.0}{(-20.0)} & \gcell{204.2}{(-38.4)} & \gcell{210.0}{(-50.0)} & \gcell{230.0}{(-130.0)} \\ \hline
        
        no specials & \cell{190.0}{(+0.0)} & \cell{200.0}{(+0.0)} & \cell{220.0}{(+0.0)} & \gcell{232.8}{(-9.7)} & \gcell{240.0}{(-20.0)} & \gcell{330.0}{(-30.0)} \\ \hline
        
        baseline & \cell{190.0}{(baseline)} & \cell{200.0}{(baseline)} & \cell{220.0}{(baseline)} & \cell{242.6}{(baseline)} & \cell{260.0}{(baseline)} & \cell{360.0}{(baseline)} \\ \hline
        
        strict & \cell{190.0}{(+0.0)} & \rcell{210.0}{(+10.0)} & \rcell{230.0}{(+10.0)} & \rcell{254.3}{(+11.7)} & \rcell{270.0}{(+10.0)} & \rcell{390.0}{(+30.0)} \\ \hline
    \end{tabular}
\captionof{table}{Condition Comparison Summary Table for the Small City (Population Cluster = 100, Trade Size = 10)}
    \label{tab:tables_small}
\end{center}

\begin{center}
    \begin{tabular}{|l|c|c|c|c|c|c|}
        \hline
        \rowcolor{headerblue}
        \headercell{Condition}{} & 
        \headercell{5\%}{(abs / diff)} & 
        \headercell{25\%}{(abs / diff)} & 
        \headercell{Median}{(abs / diff)} & 
        \headercell{Mean}{(abs / diff)} & 
        \headercell{75\%}{(abs / diff)} & 
        \headercell{95\%}{(abs / diff)} \\ \hline
        
        generous-nospecial & \cell{170.0}{(+0.0)} & \cell{180.0}{(+0.0)} & \gcell{180.0}{(-10.0)} & \gcell{183.6}{(-23.5)} & \gcell{190.0}{(-30.0)} & \gcell{200.0}{(-90.0)} \\ \hline
        
        generous-baseline & \rcell{180.0}{(+10.0)} & \cell{180.0}{(+0.0)} & \gcell{180.0}{(-10.0)} & \gcell{184.0}{(-23.1)} & \gcell{190.0}{(-30.0)} & \gcell{200.0}{(-90.0)} \\ \hline
        
        no specials & \cell{170.0}{(+0.0)} & \cell{180.0}{(+0.0)} & \cell{190.0}{(+0.0)} & \gcell{198.0}{(-9.2)} & \gcell{200.0}{(-20.0)} & \gcell{260.0}{(-30.0)} \\ \hline
        
        baseline & \cell{170.0}{(baseline)} & \cell{180.0}{(baseline)} & \cell{190.0}{(baseline)} & \cell{207.1}{(baseline)} & \cell{220.0}{(baseline)} & \cell{290.0}{(baseline)} \\ \hline
        
        strict & \cell{170.0}{(+0.0)} & \rcell{190.0}{(+10.0)} & \rcell{200.0}{(+10.0)} & \rcell{216.3}{(+9.1)} & \rcell{230.0}{(+10.0)} & \rcell{310.0}{(+20.0)} \\ \hline
    \end{tabular}
\captionof{table}{Condition Comparison Summary Table for the Large City (Population Cluster = 1000, Trade Size = 20)}
    \label{tab:tables_large}
\end{center}

\begin{figure}
    \centering
    \includegraphics[width=1\linewidth]{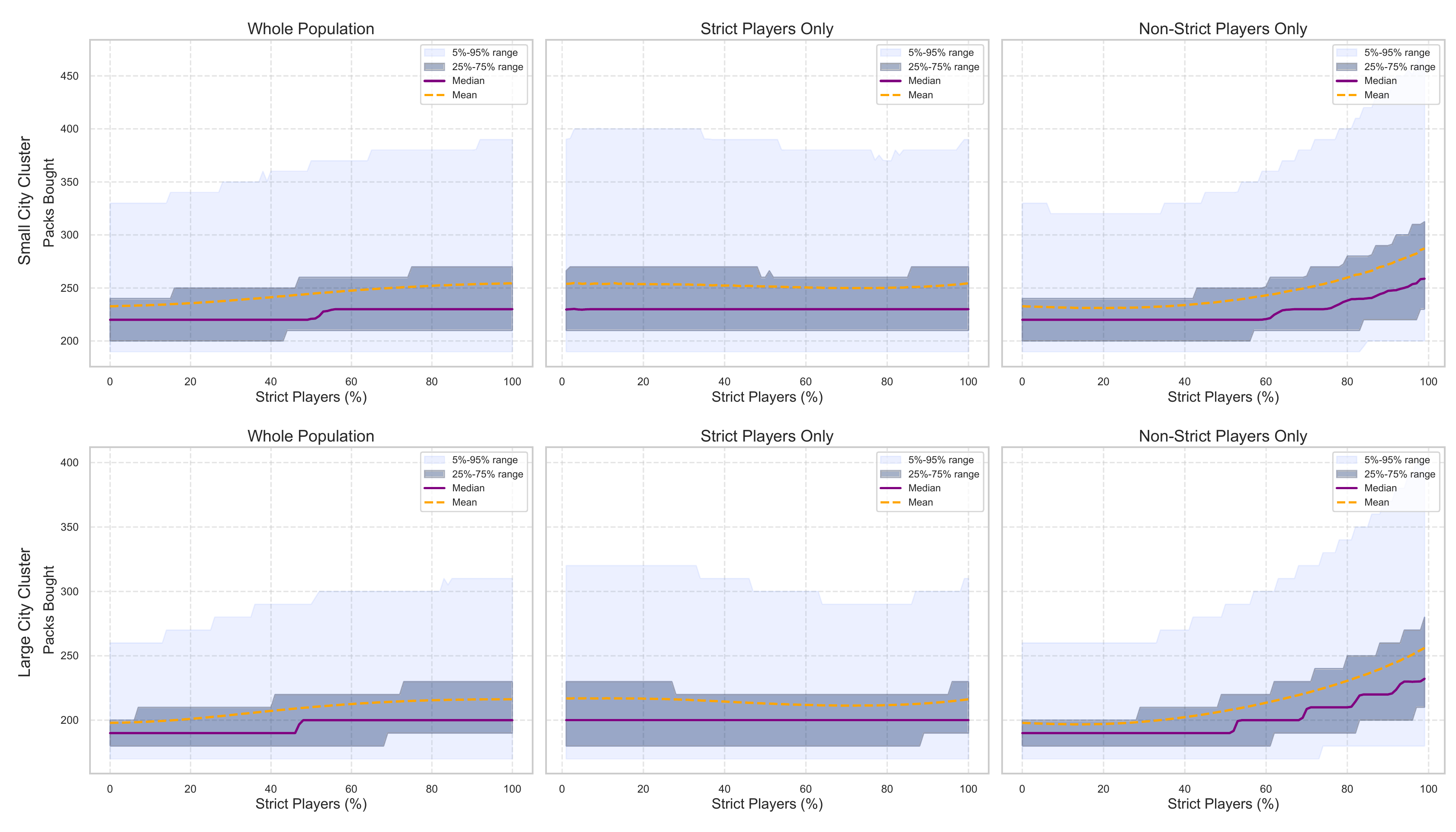}
    \caption{Impact of strict trader prevalence on album completion costs within a mixed population. The line graphs illustrate how the number of required sticker packs changes as the percentage of strict traders increases from 0\% to 100\% within a population of 1,000 players (executing 20 trades per round) operating under a baseline one-to-two exchange ratio for metallic stickers. The data is segmented into three panels: the Whole Population (left), the subset of Strict Players Only (center), and the subset of Non-Strict Players Only (right). In each panel, the solid purple line represents the median completion cost, and the dashed orange line represents the mean cost. The shaded regions denote the distribution's variance: the dark blue band captures the interquartile range (25th to 75th percentiles), while the light blue band represents the 5th to 95th percentiles, highlighting the cost extremes. The graphs demonstrate that an increasing prevalence of strict traders degrades overall network efficiency, driving up costs particularly for the non-strict minority, while the strict players maintain persistently poor completion metrics across all prevalence levels.}
    \label{fig:strict_sweep}
\end{figure}

While the strict strategy demonstrated a detrimental collective effect, we hypothesized that this impact might be mitigated if such behavior remained at a low prevalence within the population. To investigate this, we evaluated multiple hybrid scenarios by systematically sweeping the prevalence of strict traders across populations that otherwise operated under the baseline one-to-two trade rule (Figure \ref{fig:strict_sweep}). Our findings indicate that the negative impact of strict traders—particularly on the least fortunate 5\% of collectors—began to manifest even when their prevalence was below 20\%. In contrast, a noticeable shift in the median completion cost for the overall population emerged only after the prevalence of strict traders surpassed the 50\% threshold in both city configurations. Interestingly, when strict players constituted a minority of just below 50\%, they experienced worse individual outcomes than their non-strict peers. However, as strict traders became the majority, the network dynamics shifted dramatically: the environment became highly restrictive for the non-strict minority, triggering a rapid deterioration in their performance, while the overall efficiency of the strict majority remained persistently poor.

The second category of strategies addressed the inherent asymmetry in players' progression toward completing their collections. We observed that certain players nearing completion frequently adopted a generous approach when interacting with peers who required significantly more stickers. In these instances, the advanced collector would surrender their valuable duplicates to fulfill the peer's specific needs, accepting arbitrary, unneeded duplicates in return merely to balance the transaction inventory. We also noted extreme cases of altruism in the field, such as gifting a final missing sticker to a player completely lacking tradable assets, or collectors with fully completed albums continuing to trade simply to acquire aesthetically preferred items. To capture these behaviors, we synthesized these altruistic interactions into a unified generous strategy within our computational model, allowing us to systematically evaluate its overall effect on the network's collective trading dynamics.

\begin{figure}
    \centering
    \includegraphics[width=1\linewidth]{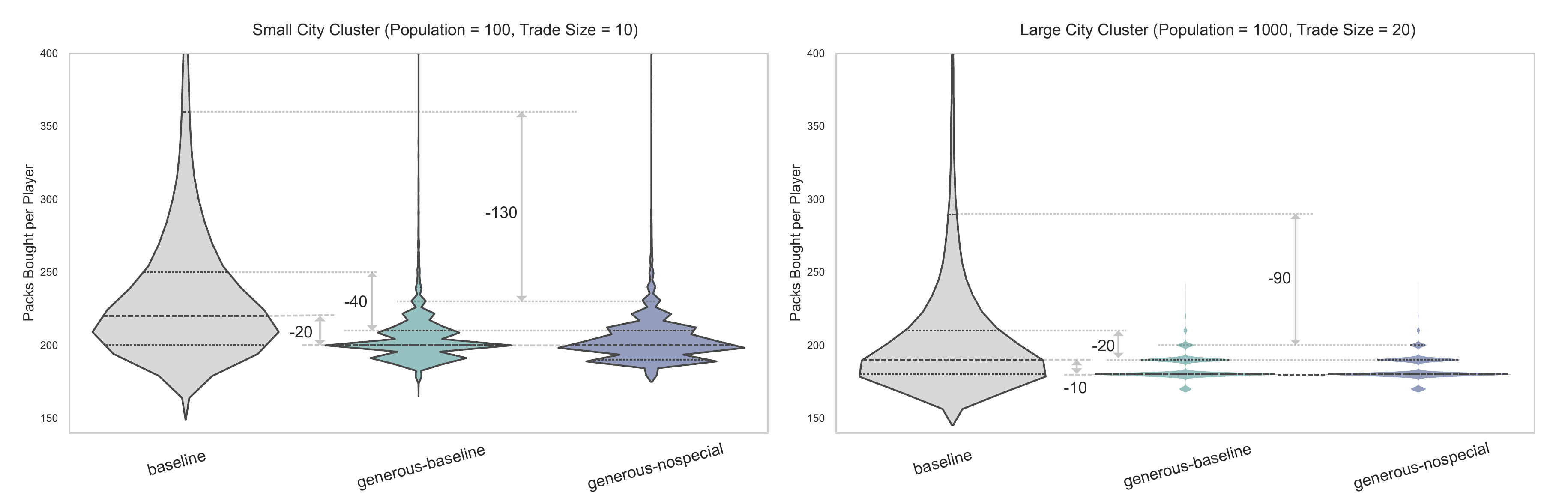}
    \caption{\textbf{Impact of generosity on album completion costs under different trading paradigms.} Violin plots depict the distribution of required sticker packs per player across three conditions: baseline (standard 1:2 special ratio, no generosity), generous-baseline (standard 1:2 special ratio with a 100\% generous population), and generous-nospecial (all stickers traded 1:1 with a 100\% generous population). The analysis contrasts two network topologies: a Small City Cluster (population = 100, trade size = 10; left panel) and a Large City Cluster (population = 1,000, trade size = 20; right panel). Horizontal dashed lines within the violins denote the 25th, 50th (median), and 75th percentiles. Dotted horizontal lines and corresponding numeric annotations indicate the absolute marginal difference in required packs at the median and the 95th percentile (the least lucky 5\% of collectors) compared to the non-generous baseline. The introduction of generosity significantly compresses the upper tail of the distribution, drastically reducing the cost burden for the least fortunate players across both network sizes. Furthermore, combining generosity with a simplified trading rule (generous-nospecial) yields the most dramatic efficiency gains and distribution homogenization, particularly in the Small City Cluster.}
    \label{fig:generosity_pop}
\end{figure}

The positive net effect of the generous strategy was consistently observable across all simulated conditions (Figure \ref{fig:generosity_pop}). Compared to the baseline, the median completion cost decreased by 10 packs in the large-city configuration and by 20 packs in the small-city configuration. The most substantial benefits were localized among the least fortunate 5th percentile of collectors, whose required purchases dropped by 90 packs in the large city and by an extraordinary 130 packs in the small city. Furthermore, examining the distribution shapes via violin plots revealed a distinct emergent phenomenon. In simulations with a high volume of trades, the outcomes showed strong structural convergence, with nearly all participants completing their albums at approximately the same cost threshold. This homogenization likely reflects a heightened functional coupling among the players' individual dynamics, indicating that widespread altruism effectively synchronizes completion rates across the entire network.

\begin{figure}
    \centering
    \includegraphics[width=1\linewidth]{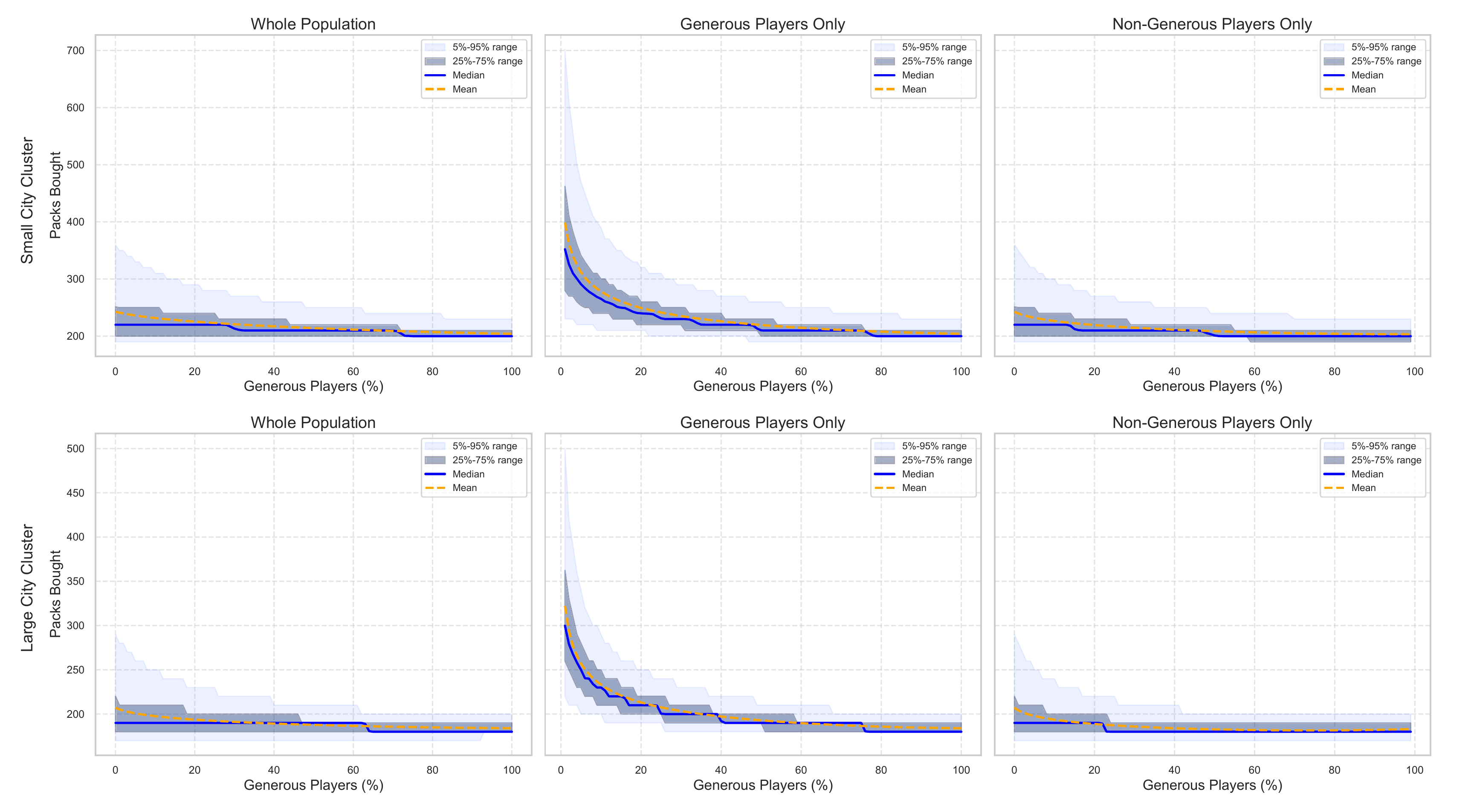}
    \caption{\textbf{Impact of generous player prevalence on album completion costs within a mixed population.} The line graphs illustrate how the number of required sticker packs changes as the percentage of generous traders increases from 0\% to 100\%. The analysis contrasts two network topologies: a Small City Cluster (population = 100, trade size = 10; top row) and a Large City Cluster (population = 1,000, trade size = 20; bottom row). For each topology, the data is segmented into three panels: the Whole Population (left), the subset of Generous Players Only (center), and the subset of Non-Generous Players Only (right). In each panel, the solid blue line represents the median completion cost, and the dashed orange line represents the mean cost. The shaded regions denote the distribution's variance: the dark blue band captures the interquartile range (25th to 75th percentiles), while the light blue band represents the 5th to 95th percentiles. The graphs demonstrate that an increasing prevalence of generous traders significantly reduces overall network costs and variance. However, when generous players are a minority (below ~50\%), they incur higher individual completion costs than their non-generous peers, a deficit that disappears as their prevalence approaches the majority.}
    \label{fig:generosity_sweep}
\end{figure}

Finally, we examined the impact of varying the prevalence of generous players within the population (Figure \ref{fig:generosity_sweep}). Increasing the proportion of generous individuals yielded an immediate reduction in the median completion cost and, notably, a substantial decrease in the overall cost variance across the network. However, an analysis of individual outcomes revealed that generous players incurred higher personal completion costs than their non-generous peers when the generous players constituted a minority. This individual performance deficit persisted until the prevalence of generous players reached approximately 50\%, at which point their completion costs finally converged with the overall population average.

\section*{Discussion}

Our simulation results offer a clear window into how localized community trading norms fundamentally dictate the efficiency of collective action within the Panini 2026 World Cup sticker economy. The data reveals a stark contrast between the systemic friction caused by overprotective behavioral rules and the massive efficiency gains unlocked by altruism. On the one hand, rigid protectionist norms, such as the strict strategy that segregates metallic stickers, trap liquidity within localized clusters. On the other hand, a generous trading strategy successfully breaks through interaction bottlenecks to mobilize stagnant assets, dramatically compressing the long tail of bad luck. Widespread generosity triggers a powerful functional coupling that effectively synchronizes completion rates across the entire network. Interestingly, both behavioral extremes exhibit unique social dynamics at a 50\% prevalence threshold: strict majorities severely penalize the non-strict minority while locking themselves into low efficiency, whereas generous minorities temporarily absorb a personal cost deficit that is completely neutralized once they become the majority. Ultimately, these findings demonstrate that the true worth of the collection is anchored in cooperation, showing that widespread altruism transforms an expensive, isolated hobby into a resilient, optimized public good. 

By mapping these empirical simulation outcomes onto formal game-theoretic and evolutionary frameworks, we can conceptualize the Panini sticker economy as a classic public goods dilemma defined by a profound structural friction between short-term individual optimization and long-term collective welfare. In a standard $n$-party public goods scenario, individual rational actions typically aggregate into Pareto-suboptimal collective outcomes because defection dominates cooperation under the classic utility inequality (T > R > P > S) \cite{Doebeli2005}. The rigid, overprotective, strict strategy observed in physical meetups precisely mirrors this noncooperative Nash equilibrium \cite{barrett_tipping_2017}; by hyper-focusing on individual protection and segregating cross-category assets, players inadvertently trigger mutual punishment (P), trapping duplicate liquidity within localized clusters and exacerbating severe interaction bottlenecks. Conversely, the altruistic generous paradigm subverts this mathematical trap by fundamentally altering how the trading network processes asymmetric progression and resource variance. 

From an evolutionary psychology perspective, the human mind possesses an evolved capacity that suppresses punitive outrage when a peer's cooperative failure is driven by accidental constraints or structural bad luck \cite{BillingsleyLosin2017,Deltonetal2012}. Because the massive cost burden concentrated at the 95th percentile of our model stems entirely from the random misfortune of poor sticker-pack draws rather than deliberate free-riding, widespread generosity serves as a highly adaptive social buffer. Generous agents willingly absorb a temporary individual performance deficit—effectively enduring the "sucker's payoff" while they remain a minority—to mobilize stagnant assets and bail out their less fortunate peers. This self-organizing behavioral matrix aligns seamlessly with Elinor Ostrom's foundational principles of common-pool resource governance \cite{Ostrom1990}, proving that informal, localized community norms can optimize the distribution of shared, subtractable assets without relying on top-down state coercion or rigid market enclosures. Once generous behavior crosses the critical 50\% prevalence threshold, this collective altruism compresses the long tail of bad luck, completely neutralizes individual cost deficits, and achieves a powerful functional coupling that transforms a highly competitive, isolated hobby into a resilient and optimized public good. 

From a pedagogical perspective, this collective endeavor functions as a vital micro-environment for childhood socialization and the practical development of essential collaborative skills \cite{Sameketal2020}. Rather than operating as an isolated or purely commercial hobby, the social architecture of sticker exchange provides young children with a low-stakes, highly interactive arena to actively practice complex behavioral competencies such as strategic negotiation, peer-seeking, and organic community chatter. More importantly, our simulation results demonstrate that this phenomenon offers an ideal educational framework for instilling interpersonal empathy and a foundational understanding of public goods. By navigating a system characterized by random pack-draw variance and structural asymmetries, young collectors are exposed to a vivid, real-world lesson in the friction between short-term individual optimization and long-term collective welfare. Engaging in an altruistic, generous strategy—in which advanced players surrender highly valued assets to assist peers with no apparent self-interest—teaches young minds that cooperating beyond narrow self-interest is not a mathematical deficit. Instead, it functions as a powerful social mechanism that dramatically compresses the long tail of bad luck for the group's least fortunate members \cite{townsend_human_2023}. Ultimately, this hobby transforms what could easily be a competitive, transaction-driven exercise into a shared ethical lesson, illustrating how localized altruism successfully turns an expensive, individualized pursuit into a resilient public good anchored in human connection and mutual reward.

While our computational model provides clear structural insights into how localized norms dictate network efficiency, it carries several inherent limitations that contrast with the significantly higher complexity of real-world collecting ecosystems. In physical trading economies, behavioral dynamics are frequently influenced by external commercial factors, such as informal secondary markets where individuals sell duplicate stickers for cash rather than executing mutual exchanges, as well as corporate marketing mechanisms like stickers with distinct color variants, artificial rarity tiers, or promotional extra stickers that fall outside the main album structure. Furthermore, our simulation assumes a relatively balanced baseline of interaction, overlooking the reality that individual players often possess highly unequal opportunities to trade based on their immediate social capital, age, or available leisure time. Real-world trading can also be far more hyper-clusterized depending on local geography, neighborhood infrastructure, and urban density, creating severe physical bottlenecks and localized asset stagnation that a stylized agent-based network cannot fully capture. While these multi-layered economic, social, and spatial variables fall outside the scope of this initial analysis, they introduce a fascinating array of emergent dynamics that make this social phenomenon an even more fertile ground for future empirical research.

\section*{Methods}

\subsection*{Experimental Design and Parameters}

We utilized agent-based computer modeling and a Monte Carlo approach to simulate and evaluate the impact of various trading rules on the cost of completing the Panini 2026 World Cup album. To accurately reflect the physical product, the simulation sets the total album size to 980 individual stickers, sold in packs of 7. Of this total, 68 stickers are designated as having a special metallic finish. The Monte Carlo simulation runs multiple iterations, looping continuously until every player in the localized environment has successfully completed their album.

\subsection*{Network Architecture}

To mirror the social reality of sticker collecting, we modeled interactions among agents using a dynamic, cluster-based contact network. The population (e.g., 100 or 1,000 players) is partitioned into smaller local communities or groups. Rather than utilizing a static network topology, social interactions are dynamically sampled during each simulation round to consistently meet a target average number of meetings per player. These transient connections are distributed between local edges, which serve as intra-group connections representing close social circles configured by default to account for 80\% of a player's interactions, and global edges, which act as inter-group connections representing casual or external trades dynamically generated at random between different subsets to fulfill the remaining interaction quota.

\subsection*{Purchasing and Trading}

The simulation operates in discrete rounds, divided into a purchasing phase and an exchange phase. In each round of the purchasing phase, any player who has not yet completed their album purchases a bulk quantity of packs, defaulting to 10 per round, with the stickers within these packs drawn at random with uniform probability from the pool of 980 available slots. Acquired stickers are logged into a player's collection, while duplicates are stored in a separate inventory for trading. To ensure fairness and simulate organic social mixing, the subsequent exchange phase features a new set of contact pairs dynamically generated each round based on the network parameters, with the sequence of these pairings randomly shuffled so that players possessing duplicates can then interact with their dynamically assigned neighbors to exchange needed stickers.

\subsection*{Trading Rules and Special Stickers}

When two connected players execute a trade, the algorithm strictly regulates the exchange of regular versus metallic stickers to reflect localized community norms. The standard exchange protocol dictates three sequential phases:
\begin{enumerate}
    \item Special-for-Special (1-to-1): Players mutually exchange needed metallic stickers at a strict one-to-one ratio.
    \item Normal-for-Normal (1-to-1): Players mutually exchange the required regular stickers at a strict 1-to-1 ratio.
    \item Special-for-Normal (1-to-2): If one player has duplicate specials and the other has duplicate normals, they trade at a one-to-two ratio, reflecting the perceived higher value of the metallic finish.
\end{enumerate}

\subsection*{Behavioral Asymmetry and Generous Players}

To evaluate the behavioral dynamics of asymmetric trading, the model introduces a "generous" trait to a designated subset of the population. In standard trades, exchanges only occur when there is mutual need. However, if a standard trade leaves player A still needing stickers from player B, and player B is designated as "generous," player B will surrender their needed duplicates to player A. To prevent inventory loss, these generous trades are "balanced" using stickers player B already possesses. The generous algorithm calculates the point value of the gifted stickers (2 points for a special, 1 point for a normal) and deducts an equivalent amount from player A's unneeded duplicates to finalize the exchange.

\subsection*{Simulation Scenarios}
To evaluate how localized community norms dictate the efficiency of collective action, we established a baseline and three comparative homogeneous simulation scenarios:
\begin{itemize}
    \item Vanilla No-Special Scenario: This condition represents a simplified environment without metallic stickers and without generous players. The only permitted trades are standard one-to-one exchanges of normal stickers based on mutual need.
    \item All Generous Scenario: Retaining the vanilla condition of no metallic stickers, this scenario sets the population to be entirely composed of generous players. Players are willing to surrender needed duplicates to assist a peer in completing their album and to accept unneeded duplicate stickers to balance the inventory exchange.
    \item Baseline Condition: This scenario introduces the 68 metallic stickers into the collection pool. Trades involve a mixed economy in which metallic stickers can be exchanged for normal stickers at a 1:2 ratio, compensating for the perceived higher value of the special finish.
    \item Strict Condition: Also featuring metallic stickers, this condition models rigid community norms in which special stickers are exchanged exclusively for other special stickers. The trade of metallic stickers for normal stickers is strictly prohibited, with exchanges limited to a one-to-one ratio.
\end{itemize}

\subsection*{Hybrid Simulation Scenarios}

In addition to the homogeneous environments, we evaluated two hybrid scenarios to model populations featuring mixed player behaviors. For both hybrid models, the population prevalence of the targeted player type was systematically varied from 0\% to 100\% in 1\% increments to analyze the gradient of their impact. The first scenario, Mixed Trading Norms, examines a population divided between two distinct trading behaviors regarding the metallic stickers, where one subset of players adheres strictly to the Strict Condition by exchanging only special-for-special stickers, while the remaining players operate under the Baseline Condition, actively engaging in special-for-normal exchanges at a one-to-two ratio. The second scenario, Mixed Generosity, evaluates a population comprising generous and non-generous players who all conduct standard trades under the Baseline Condition, using the one-to-two exchange ratio for metallic stickers, with the distinction that the generous subset is also willing to resolve asymmetric trades by surrendering needed duplicates to help peers complete their collections. 

\subsection*{Data Aggregation and Statistical Analysis}
To ensure robustness and account for network variability, each simulation condition was initialized with 64 unique random seeds to govern the dynamics of the small-world contact networks. For every initialized configuration, the Monte Carlo simulation was executed 100 times, yielding a total of 6,400 individual simulation runs per condition. For the final statistical analysis, all simulation outputs for the same condition were aggregated into a single comprehensive dataset. The evaluation of completion costs, as well as comparisons of individual and collective efficiency across different scenarios, was subsequently conducted by analyzing percentiles in this unified dataset.

\bibliography{sample}

\section*{Acknowledgements}

We thank Francisco A. Costa for his support during the fieldwork and Vitor Lopes-dos-Santos for instigating the discussions. This research was supported by the High Performance Computing Center at UFRN (NPAD/UFRN), INCT NeuroComp, CNPq, Itaú ICTi, CAPES and UFRN/PROPESQ.

\section*{Author contributions statement}

L.F.A: Conceptualization, Investigation, Writing – Review \& Editing. C.A.C.: Conceptualization, Investigation. C.R.-C.: Conceptualization, Methodology, Software, Formal Analysis, Investigation, Writing – Original Draft, Visualization.

\section*{Competing interests}

The authors declare that they have no known competing financial interests or personal relationships that could have appeared to influence the work reported in this paper.

\end{document}